\documentclass[english,twocolumn,twocolumn,showpacs,preprintnumbers,amsmath,amssymb,superscriptaddress]{revtex4}
\usepackage[T1]{fontenc}
\usepackage[latin1]{inputenc}
\usepackage{graphicx}

\makeatletter

\makeatletter

\makeatletter

\usepackage{color}

\usepackage{dcolumn}

\usepackage{bm}

\newcommand{\pt}{ p_{\rm t}}

\def\lsim{\mathrel{\rlap{\lower4pt\hbox{\hskip1pt$\sim$}}
    \raise1pt\hbox{$<$}}}         
\def\gsim{\mathrel{\rlap{\lower4pt\hbox{\hskip1pt$\sim$}}
    \raise1pt\hbox{$>$}}}         

\usepackage{babel}
\makeatother
\begin{document}

\title{
On the Role of Prompt Photons in the  Anisotropic Emission of Direct Photons

---- Direct Photons from Au+Au collisions at $\sqrt{s_{NN}}=200$~GeV with IP-Glasma Initial Condition
}

\author{Fu-Ming Liu}

\affiliation{Institute of Particle Physics, Central China Normal University, Wuhan,
China}

\email{liufm@ccnu.edu.cn}

\date{\today}

\begin{abstract}
The anisotropic emission of direct photons from Au+Au collisions at $\sqrt{s_{NN}}$=200 GeV was calculated using a (3+1)-dimensional viscous hydrodynamic model with the impact parameter Glasma initial condition. The transverse momentum spectra of direct photons in different centrality bins (0-20\%, 20-40\%, and 40-60\%) are in good agreement with experimental data measured at RHIC. 
For the elliptic flow $v_{2}$ and triangular flow $v_{3}$, the agreement is centrality dependent, showing good correspondence for the 20-40\% bin but underprediction for the 0-20\% bin and overprediction for the 40-60\% bin. After carefully accounting for the contribution from prompt photons, the measured $v_{2}$ of direct photons is no longer too large to explain. An overestimation of the prompt photon yield can suppress the $v_2$ and $v_3$ of direct photons to values lower than those observed in the experimental data.
\end{abstract}
\maketitle

\section{Introduction}
Since the discovery of a large elliptic flow for direct photons at RHIC~\cite{PHENIX2, PHENIX3, PHENIX:2015igl} and LHC~\cite{ALICE2}, accurately describing the collision system to simultaneously explain experimental data for both hadrons and photons including their yields, elliptic flow, and triangular flow, has presented a significant challenge. A highly regarded calculation of direct photon production was performed using a (3+1)-dimensional viscous hydrodynamical model with the IP-Glasma initial condition, wherein the obtained elliptic flow ($v_{2}$) and triangular flow ($v_{3}$) of direct photons were found to be lower than the experimental data~\cite{Paquet:2015lta}. Subsequent efforts have investigated various factors, such as the viscous correction to the photon emission rate~\cite{Paquet:2015lta} and the pre-equilibrium evolution of the collision system~\cite{Gale:2021emg}. 

Our group has also conducted extensive studies on direct photon production~\cite{Liu:2007tw, Liu:2008eh, Liu:2009kta,Liu:2011dk, Liu:2012ax}. To simplify the analysis, we consider two primary sources of direct photons: prompt photons and thermal photons. Prompt photons are emitted at the initial impact of the colliding heavy ions. Due to the long mean free path of photons, interactions between the emitted prompt photons and the subsequently formed hot, dense matter are negligible; consequently, prompt photons are emitted isotropically in azimuthal angle and therefore contribute zero to the elliptic flow ($v_{2}$) and triangular flow ($v_{3}$) of direct photons at any centrality. 

Thermal photons, in contrast, are emitted throughout the expansion of the hot, dense medium created in the collision. For these thermal photons, we repeat the well-established calculation using the same IP-Glasma initial condition. The expansion of the collision system is described by a (3+1)-dimensional viscous hydrodynamical model coded within the EPOS3102 framework~\cite{Werner:2013tya}. Results for thermal photons are expected to be consistent with previous results from the Paquet group\cite{Paquet:2015lta}. 

Our results for the transverse momentum ($\pt$) spectra, $v_{2}$, and $v_{3}$ of direct photons will be compared with experimental data. The dependence of the $v_{2}$ and $v_{3}$ of direct photons on event plane determination will also be examined. Because prompt photons contribute zero to the $v_{2}$ and $v_{3}$ of direct photons, the observed $v_{2}$ ($v_{3}$) of direct photons corresponds to the $v_{2}$ ($v_{3}$) of thermal photons multiplied by the percentage of thermal photons in the total direct photon yield. Therefore, the calculation of prompt photons is also critical, as it strongly influences the second factor, namely, the percentage of thermal photons in the total direct photon yield. Consequently, our calculation of prompt photons must be reliable and will also be compared with previous results\cite{Paquet:2015lta}. 

This paper is organized as follows. Section 2 presents the calculation approach for both thermal and prompt photons. In section 3, our calculation results are presented and compared with both experimental data and previous calculations. A discussion and conclusions are provided in section 4

 \section{Computational Methodology}
A significant challenge lies in constructing an accurate description of the collision system capable of simultaneously explaining experimental data for both hadrons and photons, including yields, elliptic flow, and triangle flow. The measured $v_{2}$ and $v_{3}$ of direct photons depend on the determination of collision centrality and the event plane~\cite{PHENIX:2015igl}, which are derived from the charged hadrons detected. Therefore, the calculation of direct photon observables must account for hadron production. Consequently, it is essential to provide a comprehensive picture of the space-time evolution of the entire collision system, along with the mechanisms for hadron and photon production. 

The collision process is described as a three-step sequence. 

Step 1) The two incoming heavy ions make initial contact. At this stage, the quarks and gluons within the heavy ions collide, producing prompt photons and jets. Jets are not discussed further in the following analysis. Prompt photons possess a long mean free path and escape from the bulk of the collision system.
 
 Step 2) The bulk matter undergoes a pre-equilibrium evolution and thermalizes at the initial time. Photon production is neglected during this stage for two reasons. First, the photon yield is negligible, being significantly lower than that of prompt photons at high $\pt$ and substantially lower than that of thermal photons at low $\pt$~\cite{Gale:2021emg}. Second, the flow velocity of the bulk is small at this stage and cannot generate a significant anisotropy in photon emission, rendering it irrelevant to the observed large anisotropic emission of direct photons. 
 
 The distribution of energy and momentum at the initial time defines the initial condition (IC), which can be determined by either the distribution of constituent particles, i.e., EPOS IC, or the distribution of the electromagnetic (color) field excited by these particles, i.e., IP-glasma IC. In this work, we employ the IP-glasma IC~\cite{Schenke:2012wb, Schenke:2012hg} for the event-by-event hydrodynamical evolution of the hot dense matter formed in AuAu collisions at $\sqrt{s_{NN}}$=200 GeV, with an initial time of $\tau_0$=0.4 fm/c.
 
 Step 3) The bulk, a hot and dense matter, is assumed to undergo two distinct phases: the quark-gluon plasma (QGP) and the hadronic gas (HG). The collective motion of this bulk matter is described as that of a viscous fluid governed by hydrodynamics. Thermal photons are emitted from the bulk during both the QGP and HG phases. As the bulk expands, it cools, and its internal interactions weaken. Subsequently, hadrons freeze out and, if within the detector's acceptance window, are registered by the detector. In this work, we employ a 3+1D viscous hydrodynamic approach using the EPOS3102 framework~\cite{Werner:2013tya}, which solves the equations of relativistic viscous hydrodynamics and incorporates the mechanism for hadron freeze-out. The hydrodynamic solution provides the energy density and flow velocity for each cell in the bulk at any given time. Furthermore, the temperature of each cell is determined using the equation of state implemented in EPOS3102.

 \subsection{The transverse momentum spectrum of Prompt Photons }

Prompt photons are emitted during the initial interaction of incoming heavy ions. 
The calculation is performed to next-to-leading order for cold nuclear collisions:
\begin{eqnarray}
\frac{dN^{{\rm pr}}}{dyd^{2}\pt} & = & T_{AB}(b)\sum_{{\displaystyle ab}}\int dx_{a}dx_{b}G_{a/A}(x_{a},M^{2})G_{b/B}(x_{b},M^{2})\nonumber \\
 & \times & \frac{\hat{s}}{\pi}\delta(\hat{s}+\hat{t}+\hat{u})[\frac{d\sigma}{d\hat{t}}(ab\rightarrow\gamma+X)\\
 & + & K\sum_{c}\frac{d\sigma}{d\hat{t}}(ab\rightarrow cd)\int dz_{c}\frac{1}{z_{c}^{2}}D_{\gamma/c}(z_{c},Q^{2})],\nonumber 
\label{eq:prompt-pt}
\end{eqnarray}
where $T_{AB}(b)$ is the nuclear overlap function at an impact parameter $b$ for each centrality,  and
$G_{a/A}(x_{a},M^{2})$ and $G_{b/B}(x_{b},M^{2})$ are the parton distribution functions in nuclei A and B. 
We use the MRST 2001 leading-order parton distributions~\cite{Martin2002}. 

 Nuclear shadowing and EMC effects are accounted for using the EKS98 scale-dependent nuclear ratios $R^{\rm EKS}_a (x,A)$ \cite{Eskola1999}. 
Isospin of a nucleus with mass $A$, neutron number $N$, and proton number 
$Z$ is corrected as follows:
\begin{equation}
G_{a/A}(x) = [ \frac{N}{A} G_{a/N}(x)  
 +\frac{Z}{A} G_{a/P}(x)  ]  R^{\rm EKS}_a (x,A)  
\label{eq:EKS}
\end{equation}
The elementary processes $ab\rightarrow\gamma+X$ includes Compton scattering $qg \rightarrow\gamma q$ and annihilation  $ q\bar q \rightarrow g\gamma$.
 The elementary cross sections for the processes $ab\rightarrow\gamma+X$  and  
$ab \rightarrow cd$ are detailed in ~\cite{Owens1987}.
The photon fragmentation function $ D_{\gamma/c}(z_{c},Q^{2})$ represents 
the probability of obtaining a photon from a parton $c$ that carries a fraction $z$ of the parton's momentum.
The effective fragmentation functions for producing photons from partons can be calculated perturbatively. We employ the parameterized solutions provided by Owens~\cite{Owens1987}.
We set the factorization scale $M$ and renormalization scale $Q$ to be $M = Q = \pt$, and  $K=2$ to count high order contribution to parton $c$ production.
In contrast to \cite{Liu:2008eh}, the present work neglects the energy loss in fragmentation functions to compensate for the contribution from jet photon conversion. The calculation of the $\pt$ spectrum of prompt photons has been validated over a wide range of $\pt$ and collision energies~\cite{Liu:2008eh,Liu:2011dk}.

The equation 
\begin{equation}
\frac{dN^{{\rm pr}}}{dyd^{2}\pt} = N_{\rm coll} \frac{dN ^{NN\rightarrow \gamma} }{dyd^{2}\pt }
\end{equation}  
separates the collision centrality dependence, carried by the factor $N_{\rm coll}$, from the distribution functions and cross section information contained in ${dN ^{NN\rightarrow \gamma} }/{dyd^{2}\pt }$
For Au+Au at collisions at  $\sqrt{s_{NN}}=200$~GeV,  the value of 
$ N_{\rm coll}$  are  810.89, 213.82 and 44.38 for the 0-20\%, 20-40\% and 40-60\% centrality classes, respectively. 
To facilitate comparison, we present the quantity $ X={dN ^{NN\rightarrow \gamma} }/{dyd^{2}\pt }$ 
as a function of $\pt$ in Table~\\ref{cent}.

\begin{table}

\caption{\label{cent}
$ X= {dN ^{NN\rightarrow \gamma} }/{dyd^{2}\pt }$ as a function of $\pt$ at  $\sqrt s$=200~GeV for y=0.  }
\[
\begin{array}{cc}
     \pt ({\rm GeV/c})   &    X \\ 
           1  &   1.00683983E-05\\ 
           2  &   2.28681324E-06\\
           3  &   4.26288437E-07\\ 
           4  &   1.11610014E-07\\
           5  &   3.58652699E-08\\
           6  &   1.33102382E-08\\ 
           7  &   5.51087043E-09\\
           8  &   2.48294540E-09\\
           9  &   1.19624732E-09\\
          10  &   6.09683859E-10   \end{array}\]
\end{table}

\subsection{The transverse momentum spectrum of Thermal Photons }

The $\pt$ spectrum of thermal photons is given by 
\begin{equation}
\frac{dN^{{\rm th}}}{dyd^{2}\pt}=\int d^{4}x\Gamma(E^{*},T)\label{eq:thermal-pt}
\end{equation}
 where $\Gamma(E^{*},T)$ is the photon emission rate at temperature
$T$ and $E^{*}=p^{\mu}u_{\mu}$, $p^{\mu}$ is the four-momentum
of a photon in the lab frame and $u_{\mu}$ is the flow velocity.

For each heavy ion collision,  the hydrodynamical evolution of the bulk medium provides us the energy density $\epsilon$ 
and flow velocity $u_{\mu}$ for  each cell of the system from an  initial time $\tau_0$=0.4~fm/c  until  freeze-out. 
 We calculate the thermal photons emitted from each cell of the hot dense matter when its energy density exceeds 0.08~GeV/fm$^3$ and its temperature is greater than 0.02GeV. 
 The emitted photons carry information about the flow velocity $u_{\mu}$, which generates the elliptic flow $v_2$ and triangular flow $v_3$ of the thermal photons.
 
 A systematic study of the thermal photon emission rate $\Gamma(E^{*},T)$ was presented in Ref.~\cite{Liu:2007tw}, including contributions from both the quark-gluon plasma (QGP) and the hadronic gas (HG). This work demonstrated the competition between photon emission rates in the QGP and HG phases, as well as among various parameterized rates corresponding to microscopic processes in the HG phase, such as $\pi+\rho\rightarrow\pi+\gamma$, $\pi+\pi\rightarrow\rho+\gamma$, and $\pi+K^{*}\rightarrow K+\gamma$. In this work, we adopt the AMY rate for thermal photons from the QGP phase and the parameterized HG rates without form factors, consistent with our previous studies~\cite{Liu:2008eh, Liu:2009kta,Liu:2011dk, Liu:2012ax}.
 
 \subsection{The  elliptic flow $v_{2}$ and the triangle flow  $v_{3}$   of  Photons }

The elliptic flow $v_{2}$ and triangular flow $v_{3}$ of photons are calculated as follows.
For each simulated event, the azimuthal dependence of thermal photons is characterized by Fourier coefficients 
$ v^{\gamma}_n (\pt) $ and corresponding angles $\Phi^{\gamma}_n (\pt)$ at a given $\pt$ window and for rapidity $y=0$, according to 
\begin{eqnarray} 
v^{\gamma}_n(\pt) {\rm e}^{{\rm i}n\Phi^{\gamma}_n (\pt) } =
\frac {\int d\phi[{dN}/{dyd^{2}\pt}]{\rm e}^{{\rm i}n\phi }}{\int d\phi[{dN}/{dyd^{2}\pt}]}, 
\label{eq:decompose} 
\end{eqnarray} 
where $ [{dN}/{dyd^{2}\pt}] $ is the momentum distribution of thermal photons in a single event and $\phi$ is the azimuthal angle of the photon's momentum with respect to the reaction plane, which contains the impact parameter and beam axis. This procedure yields the $n$th-order flow harmonic coefficient $ v^{\gamma}_n (\pt) $ and the corresponding flow angle $\Phi^{\gamma}_n (\pt)$ for thermal photons relative to the reaction plane. Similarly, Eq.~\ref{eq:decompose} can be used to obtain the flow harmonics $ v^{\rm h }_n (\pt) $ and angles $\Phi^{\rm h}_n (\pt)$ for any hadron species {h}.

In hydrodynamical simulations, the orientation of each collision system and the reaction plane are known.    
However, experimentalists must reconstruct the reaction plane on a event-by-event  basis from the detected particles. 
This reconstructed plane is referred to as the event plane, with respect to which   the  $v_{2}$ and   $v_{3}$   of  direct photons  are measured.  
The $v^{n}$ of thermal photons, $v^{\rm th}_{n}$, and direct photons, $v^{\rm dir}_{n}$, are calculated with respect to the event plane. 

To determine the event plane angle $\Phi^{\rm h}_n$, we use hadrons emitted during freeze-out from the bulk medium. Details regarding freeze-out (hadronization) can be found in \cite{Werner:2013tya}. The results for thermal photons are expected to be consistent with previous findings by the Paquet group \cite{Paquet:2015lta}.

In the PHENIX experiment, the event plane is determined using all detected charged particles, where RxNO and RxNI have rapidity windows of $-2.8< \eta < -1$ and $ 1< \eta <2.8$, respectively \cite{PHENIX:2015igl}. In contrast, hadrons from mid-rapidity are used to determine the event plane in Ref.~\cite{Paquet:2015lta}. For our calculation, we consider two rapidity options for the hadrons ($-5 \leq \eta \leq 5$ or $-1 \leq \eta \leq 1$) with $\pt > 0.3$~GeV/c to obtain $\Phi^{\rm h}_n$ and $v^{\rm h}_n$ via Eq.~\ref{eq:decompose}.

The $v_{n}$ of thermal photons  with respect to the event plane are obtained using the scalar product method. 
\begin{eqnarray}
v^{\rm th}_{n} (\pt)=\frac
{  <      v^ {\gamma}_n (\pt)      v^{\rm h}_n   {\rm cos} { n  [  \phi ^{\gamma} ( \pt)  -   \Phi^{\rm h}_n    ]    }              > }
{ \sqrt { <  (  v^{\rm h}_n ) ^{2}   >   }    },      
\label{eq:vn-EventPlane}
\end{eqnarray}
where <...> denotes an average over all events in the sample. We simulated 4000 events for each centrality class. 

Since the prompt photons have zero $v_n$,  the flow of direct photons with respect to the event plane is calculated as 
\begin{equation}
 v^{\rm dir}_n  =  \frac  {N^{\rm th}} { N^{\rm th} +N^{\rm pr} }    * v^{\rm th}_{n}  
\label{normalize}
 \end{equation}   
where 
 $ N^{\rm th} $ and  $N^{\rm pr} $ represent  ${dN^{{\rm pr}}}/{dyd^{2}\pt} $ in Eq.~\ref{eq:prompt-pt}    and 
 ${dN^{{\rm th}}}/{dyd^{2}\pt}$ in  Eq.~\ref{eq:thermal-pt}, respectively.

\section{Results} 

\begin{figure*}
\includegraphics[scale=0.6]{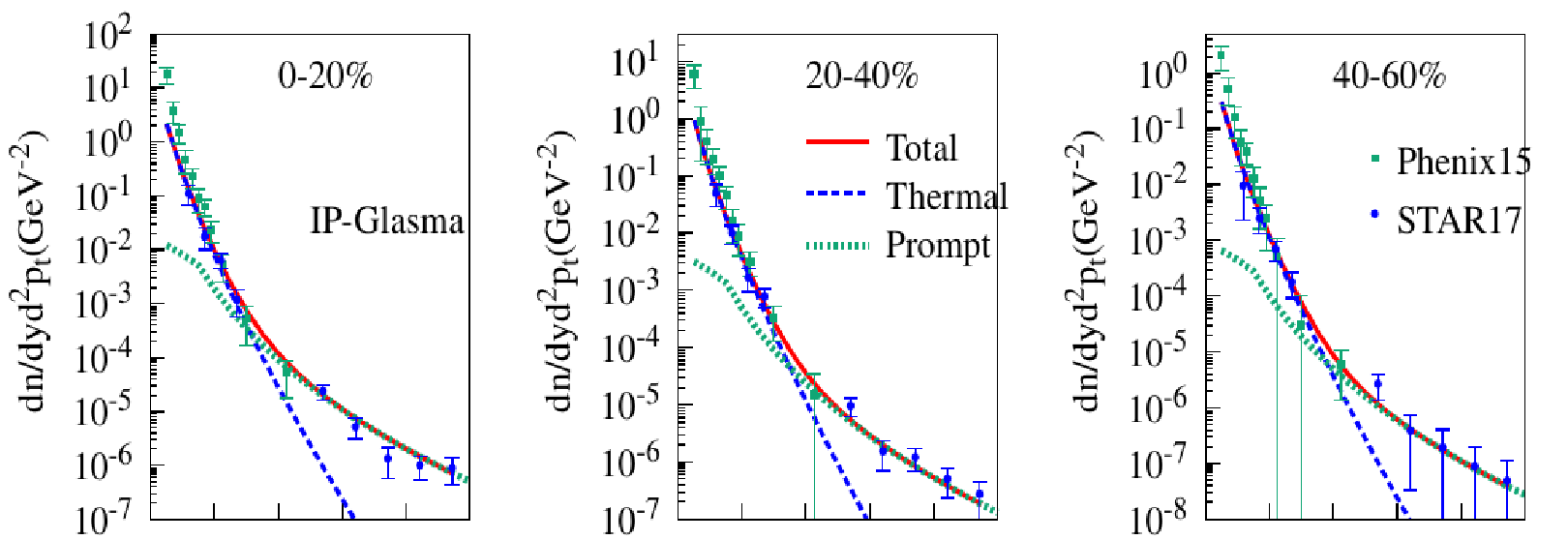}
\includegraphics[scale=0.6]{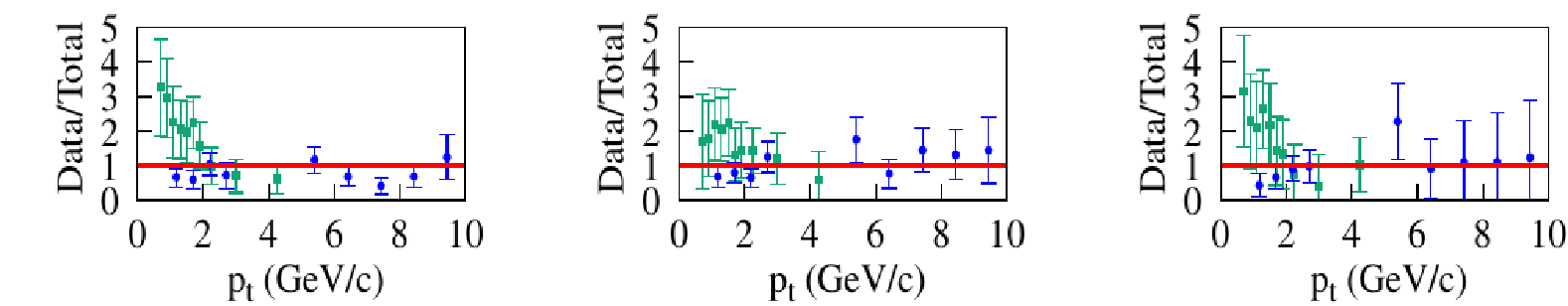}

 \caption{\label{200spectra} (Color Online) 
The calculated transverse momentum spectra of 
prompt photons (dotted lines), thermal photons (dashed lines), and total direct photons (solid lines) 
from AuAu collisions at $\sqrt{s_{NN}}=200$~GeV for centrality classes 0-20\%, 20-40\%, and 40-60\% 
are compared with data from STAR~\cite{STAR:2016use} and PHENIX~\cite{PHENIX:2015igl}. }
\end{figure*}

The calculated transverse momentum  spectra of three centrality classes are compared with  
STAR~\cite{STAR:2016use} and PHENIX~\cite{PHENIX:2015igl}  data
in  Fig.~\ref{200spectra}, where 
prompt photons, thermal photons, and total direct photons are represented by dotted, dashed, and solid lines, respectively. The upper panels clearly show the dominance of the thermal contribution at low $\pt$ and of prompt photons at high $\pt$. The ratios of Data/Total shown in the lower panel reveal that the calculated $\pt$ spectra of direct photons are in good agreement with experimental data over a wide $\pt$ range for all centrality classes. In particular, the good agreement at high $\pt$ demonstrates the reliability of the prompt photon calculations.

A comparison of our results with the widely recognized findings from the Paquet group is presented in Fig.~4 of \cite{Paquet:2015lta}. For thermal photons, the spectra from both calculations are in close agreement, as anticipated given the identical initial conditions and comparable hydrodynamic evolutions. Minor discrepancies may be attributed to the corrections applied to the photon emission rates in \cite{Paquet:2015lta}. In contrast, a significant difference is observed for prompt photons, where the value reported by Paquet et al. is approximately an order of magnitude larger than our result at $\pt=1$ GeV/c and a factor of four greater at $\pt=2$ GeV/c. This disparity can substantially influence the $v_2$ and $v_3$ of direct photons, as indicated by Eq.~\ref{normalize}.

\begin{figure*}
\includegraphics[scale=0.6]{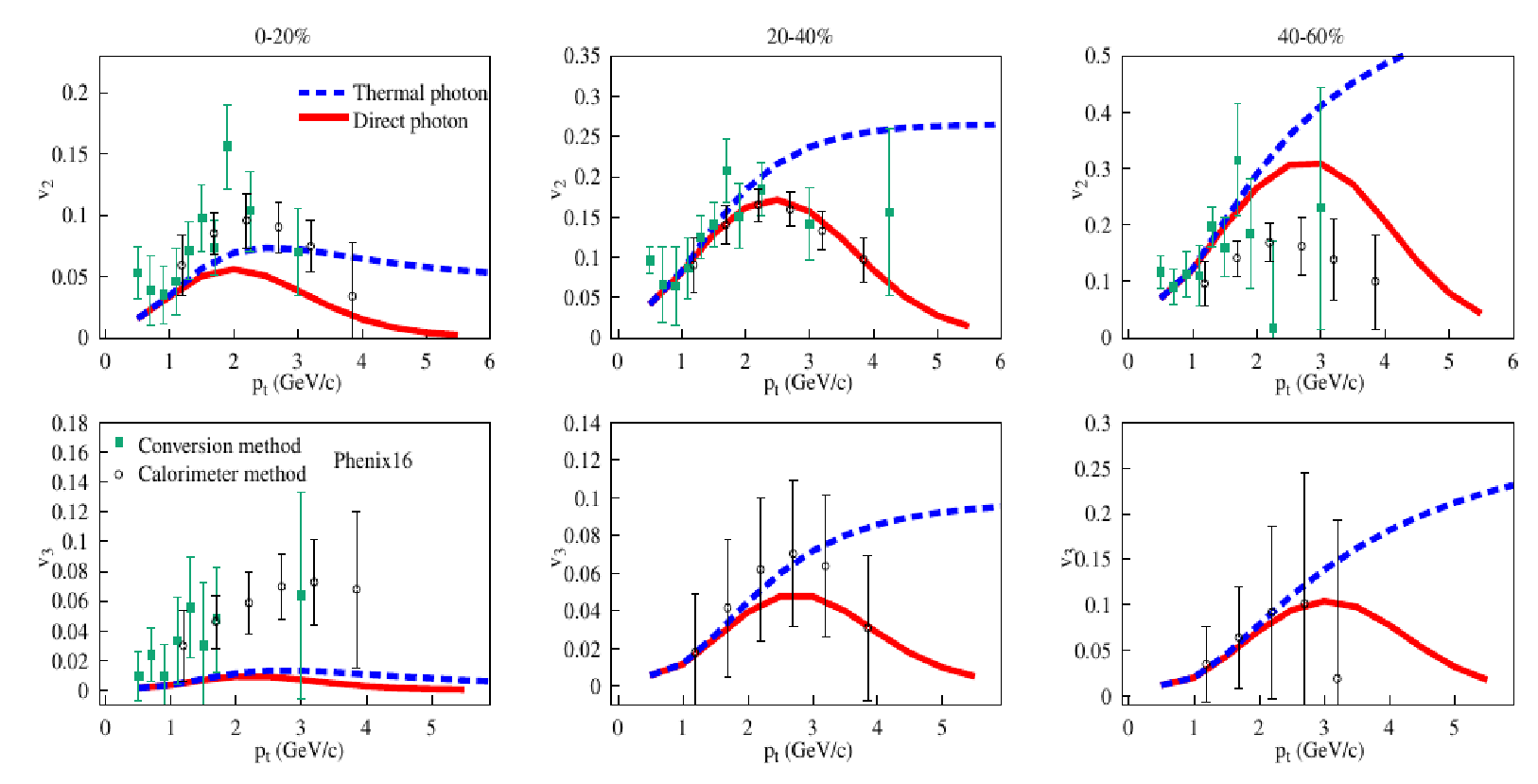}
 \caption{\label{v2v3} (Color Online) 
 The  calculated  elliptic flow $v_2$ (upper panels) and triangular flow $v_3$ (lower panels) of direct photons (solid lines) and thermal photons (dashed lines) from AuAu collisions at $\sqrt{s_{NN}}=200$~GeV for centrality class 0-20\% , 20-40\% and 40-60\%  are compared with PHENIX data~\cite{PHENIX:2015igl} (squares) .  The rapidity window  for event plane determination is $-5\leq \eta \leq 5$.   }
\end{figure*}

The calculated elliptic flow $v_2$ (upper panels) and triangular flow $v_3$ (lower panels) of direct photons (solid lines) and thermal photons (dashed lines) from AuAu collisions at $\sqrt{s_{NN}}=200$~GeV for centrality classes 0-20\%, 20-40\%, and 40-60\% are compared with PHENIX data \cite{PHENIX:2015igl} (squares) in Fig.~\ref{v2v3}. The rapidity window used for event plane determination is $-5 \leq \eta \leq 5$.
 
The calculated elliptic flow $v_2$ of direct photons (red solid lines) is
comparable to experimental data. For centrality 20-40\%, the red solid curve  passes through the data points accurately. The curve is higher than the data  for centrality 40-60\%, and lower than  the data for centrality 0-20\%.  No adjustments have been made to fit the data points. Furthermore, the measured elliptic flow of direct photons is no longer excessively large to be unexplained.

The calculated triangle flow $v_3$ of direct photons (red solid lines) is in good agreement with  the data for 20-40\% and 40-60\% centrality, whereas it underestimates the data for 0-20\%. Overall, the theoretical results for the $v_2$ and $v_3$ of direct photons are consistent with the experimental data.
We again compare our results with those of Paquet~\cite{Paquet:2015lta}, shown in Fig.~6 therein. First, we examine the elliptic flow $v_2$ of thermal photons. The thermal results from the two calculations are consistent for $\pt < 2$ GeV/c, as expected. However, a notable difference exists: our thermal $v_2$ does not decrease at high $\pt$, which may be attributed to the non-zero initial flow velocity (which is larger in more peripheral collisions).

A comparison of the elliptic flow $v_2$ for direct photons reveals that our results for the 20-40\% centrality class are evidently higher than those of Paquet. The reason for this discrepancy is readily apparent. In our calculation, the dominance of thermal photons up to $\pt=4$~GeV/c in the $\pt$ spectra (Fig.~\ref{200spectra}) causes the $v_2$ of direct photons to closely follow the thermal curves in the low $\pt$ region, $\pt<2$~GeV/c. This situation does not occur in Paquet's analysis, where the prompt curve crosses the thermal one twice and the $v_2$ of direct photons is evidently lower than the thermal curves even at low $\pt$.

\begin{figure*}
\includegraphics[scale=0.6]{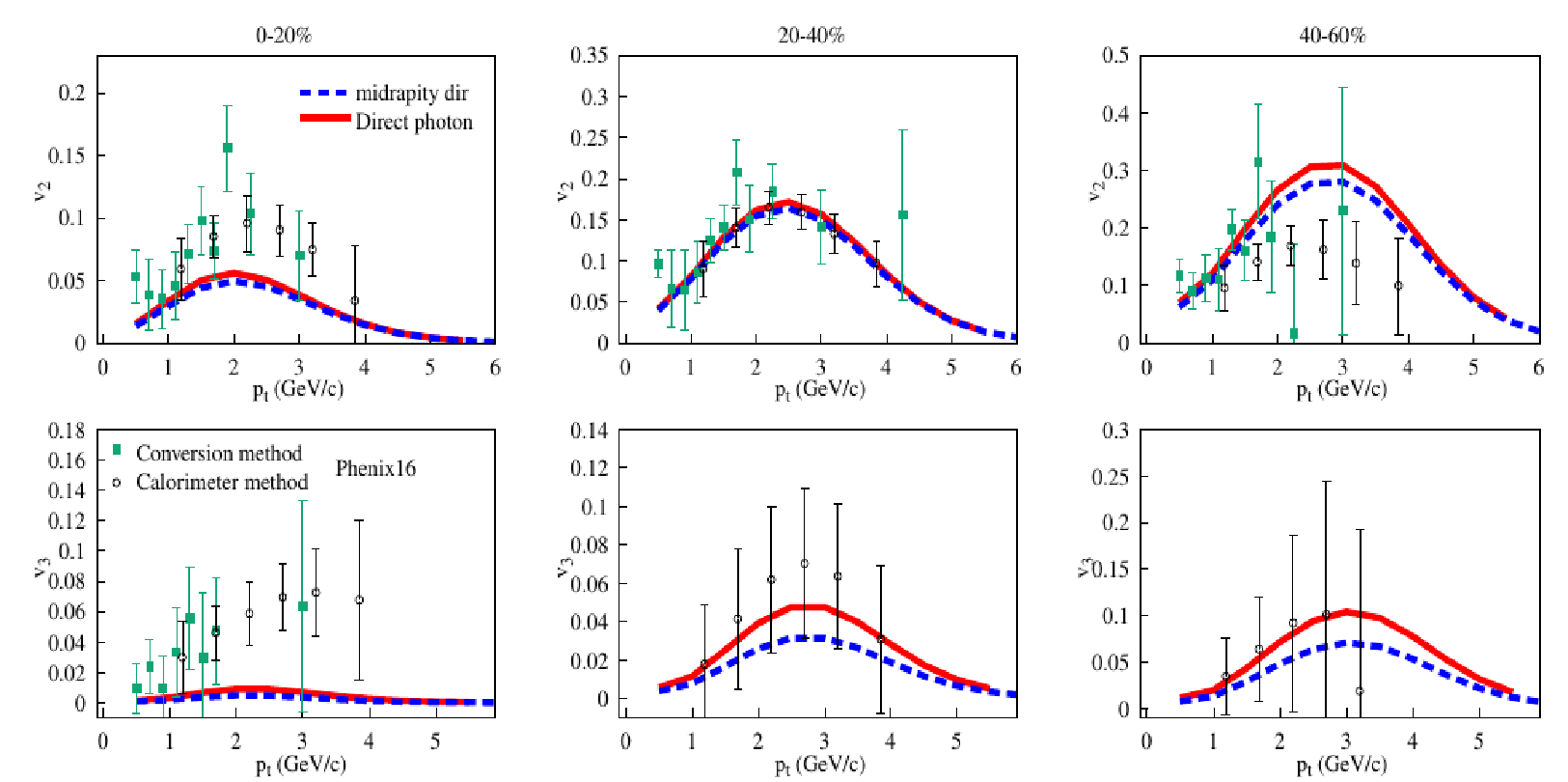}
 \caption{\label{v2v3IPG1} (Color Online)  As in Fig.~\ref{v2v3}, except that the dashed lines represent the results for direct photons within the rapidity window $-1\leq \eta \leq 1$ used for event plane determination. }
\end{figure*}

Following the methodology of \cite{Paquet:2015lta}, where the event plane is determined using midrapidity particles, Fig.~\ref{v2v3IPG1} shows the effect of the rapidity window on the elliptic flow $v_2$ and triangular flow $v_3$ of direct photons. The solid lines correspond to the results from Fig.~\ref{v2v3}, obtained with a rapidity window of $-5 \leq \eta \leq 5$, while the dashed lines represent the results for a narrower window of $-1\leq \eta \leq 1$. The influence of the rapidity window becomes more pronounced in peripheral collisions characterized by low multiplicity. Furthermore, the rapidity window has a stronger effect on $v_3$ than on $v_2$. A wider rapidity window encompasses a greater number of final-state particles, which leads to an event plane that more accurately represents the reaction plane.

\section{Conclusion}
In this paper, we briefly review the main sources of direct photon production in heavy ion collisions. Based on this simple model and without additional corrections, a satisfactory explanation is obtained for the measured $\pt$ spectra across all centrality classes. Simultaneously, the theoretical results for the $v_2$ and $v_3$ of direct photons are consistent with experimental data.

We compared our calculations with the widely recognized results of Paquet et al., finding that their approach overestimates the contribution of prompt photons by a significant factor (dependent on $\pt$), which leads to a strong suppression of the $v_2$ and $v_3$ of direct photons and yields values that are lower than the experimental data.

These findings demonstrate the significant role of prompt photons in the anisotropic emission of direct photons. Consequently, the measured elliptic flow of direct photons is no longer anomalously large and can be explained. This framework allows for a simultaneous explanation of the $\pt$ spectrum, $v_2$, and $v_3$ of direct photons.

\begin{acknowledgments}
This work was supported by the Natural Science Foundation of China (Project No. 11275081) and the Program for New Century Excellent Talents in University (NCET). The hydrodynamical simulations were supported by the Subatech laboratory.
\end{acknowledgments}

\end{document}